# What is Realism in Quantum Mechanics and How Can it be Nonlocal

*Sofia D. Wechsler*

**Abstract**

The concept of "realism" in quantum mechanics means that results of measurement are caused by physical variables, hidden or observable. Local hidden variables were proved unable to explain results of measurements on entangled particles tested far away from one another. Then, some physicists embraced the idea of nonlocal hidden variables. The present article proves that this idea is problematic, that it runs into an impasse vis-à-vis the special relativity.

**Abbreviations:**

| | |
|---|---|
| LHV | = local hidden variable |
| LHS | = left hand side |
| NLHV | = nonlocal hidden variable |
| QM | = quantum mechanics |
| RHS | = right hand side |
| SG | = Stern-Gerlach |
| w-f | = wave-function |
| w-p | = wave-packet |

## 1. Introduction

The concept of "realism" in quantum mechanics means that results of measurement of quantum objects are caused by physical variables, hidden or observable, possessed by the measured object or by the environment, and independent of the type of measurement performed. Local hidden variables (LHV) were proved by J. Bell [1] unable to explain results of measurements on entangled particles tested far away from one another[1] (see also [2], [3]).Then, there came contextual experiments that also ruled out the LHV [5], [6], [7] (see also references inside), [8]. Another argument against LHV was brought by experiments in which moving frames of coordinates are involved [9] – the LHV were proved incompatible with the special relativity. On the other hand, experiments were performed seeking frames of coordinates in which the correlations appearing in entanglements would be violated [10], [11], [12]. No such frames were found.

Vis-à-vis this situation, the idea of nonlocal hidden variables (NLHVs) began to gain ground. It is a fuzzy idea since nobody knows physical properties fitted to each wave-function (w-f) in particular, and that extend over all the space. Therefore, as in the case of the LHV, the possibility of existence of the NLHV has to be tested first of all, mathematically, and vis-à-vis all the other laws of the physics. Mathematical and physical arguments against NLHV were brought in [13] by Leggett, in base of the violation of a particular criterion, and then by N. Gisin [14], who showed that a theory of NLHV should accept that results of present measurements should depend on future events, independent of the present ones, e.g. choices of futures types of tests. The dependence on future events, contrary to the causality, seemed to the Gisin impossible.

---

[1] The present author proved that the violation of Bell-type inequalities is not enough an argument against the locality, [4], since in the inequalities Bell used classical probabilities real and positive, while in the quantum formalism are used complex amplitudes of probabilities. The latter type of calculus cannot be mapped on the former.

Though, so is the behavior of entangled particles. This fact appears implicitly in the *free will theorem* of Conway and Kochen [15], [16], the free will referring to the liberty of experimenters to choose which types of measurements to do on the particles they test.
A whole interpretation of the quantum mechanics (QM) was constructed by L. De Broglie, and continued by D. Bohm. It incorporates the hypothesis of nonlocal behavior of quantum objects [17], [18], [19]. However, it doesn't solve the problem of the dependence of the future. It also failed to cope with the relativity as explained in [20], and with the entanglements as explained in [21] section 3.

The present text also contains a proof against NLHV, which, to the difference from [14], accepts the possibility that results of present measurements depend on future events. The proof works with a type of entanglement less used in the literature, the singlet of spin 1 bosons, and uses relativistic arguments.

The next sections have the following general line: section 2 presents a couple of properties of the spin 1 bosons and of their singlet state. Section 3 describes an experiment with the spin singlet and obtains the quantum predictions for the results when moving frames are considered. Section 4 introduces the hypothesis of NLHV and shows that this hypothesis leads to a contradiction. Section 5 contains conclusions.

## 2. A couple of properties of the singlet of spin 1 bosons

While the spin projection $S_Q$ of a spin 1 boson takes on an arbitrary direction **Q** in space three values, 1, 0, or –1 (in $\hbar$ units), the square of the spin-projection, $S_Q^2$, can take only two values 1, or 0. One single eigenvector, $|0\rangle_Q$, corresponds to the eigenvalue 0, while to the eigenvalue 1 correspond two eigenvectors, denoted in this text by $|1\rangle_Q$, and $|1'\rangle_Q$ – see the expressions of these vectors in (A1) in the appendix, . These three eigenvectors form a base {Q} in the space of eigenvectors of the square of the spin 1 projections.
The square of the spin 1 projections can be measured with two Stern-Gerlach (SG) devices as in the figure 1.

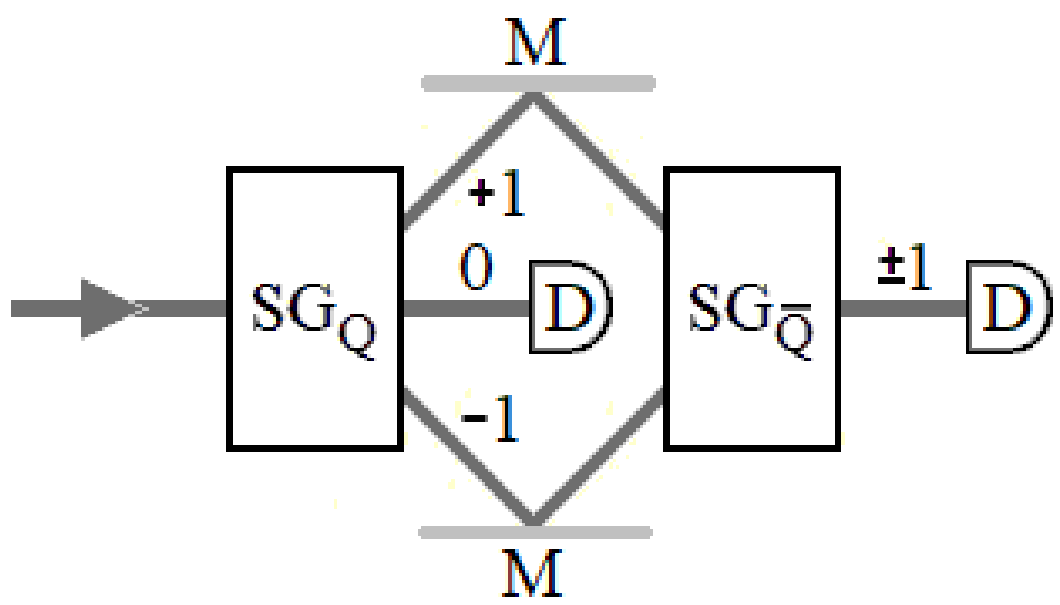

Figure 1. Measurement of the observable $S_Q^2$.

The SG device $SG_Q$ splits an incoming boson beam of spin 1 into three copies with $S_Q = 1, 0, -1$, respectively. On the beam with $S_Q = 0$ is placed a detector D, which therefore detects if $S_Q^2 = 0$. The beams with $S_Q = 1$ and $S_Q = -1$ are deflected by mirrors M toward a second SG device, $SG_{\bar{Q}}$, having the magnetic field oppositely oriented than the one in $SG_Q$. This device merges these two beams into the beam labeled $\pm$. Therefore, a detector placed on this beam detects if $S_Q^2 = 1$.

The detector D on the beam with the notation '0' detects the value $S_Q^2 = 0$, and the detector on the beam with the notation $\pm$ detects the value $S_Q^2 = 1$.

The following properties of the spin 1 boson and of their singlet state will be relevant in continuation:

**A)** Let **X**, **Y**, **Z**, be three directions in space mutually orthogonal two by two – figure 2. It is known that, to the difference of the simple projection operators $\hat{S}_X$, $\hat{S}_Y$, and $\hat{S}_Z$, the square projection operators $\hat{S}_X^2$, $\hat{S}_Y^2$, $\hat{S}_Z^2$, commute two by two. Therefore, these operators can be measured in whichever order in a single trial of the experiment, on the same particle. As the total spin of such a boson is equal to 2, the results obtained for $\hat{S}_X^2$, $\hat{S}_Y^2$, and $\hat{S}_Z^2$ are twice 1 and once 0. This is the well-known '101' law for spin 1 bosons.
The immediate consequence is that for two mutually perpendicular directions **P** and **Q** with $\mathbf{P} \perp \mathbf{Q}$, one cannot get both $S_P^2 = 0$ and $S_Q^2 = 0$,

$$\text{If } (\mathbf{P} \perp \mathbf{Q} \text{ and } S_Q^2 = 0)\text{, then, } S_P^2 = 1. \tag{1}$$

**B)** The singlet state of these bosons has the following expression in the base {Q}

$$|\psi\rangle = \frac{1}{\sqrt{3}}\left(|1\rangle_Q|1\rangle_Q - |0\rangle_Q|0\rangle_Q - |1'\rangle_Q|1'\rangle_Q\right). \tag{2}$$

One can check by introducing the expressions of the vectors of the base {Q} – see in the appendix the expressions in (A1) – that this form is invariant at a change of the direction **Q**.
From (2) results that at a measurement of the square of the spin projection in an arbitrary base {Q}, the same for the two bosons, they give the same response.

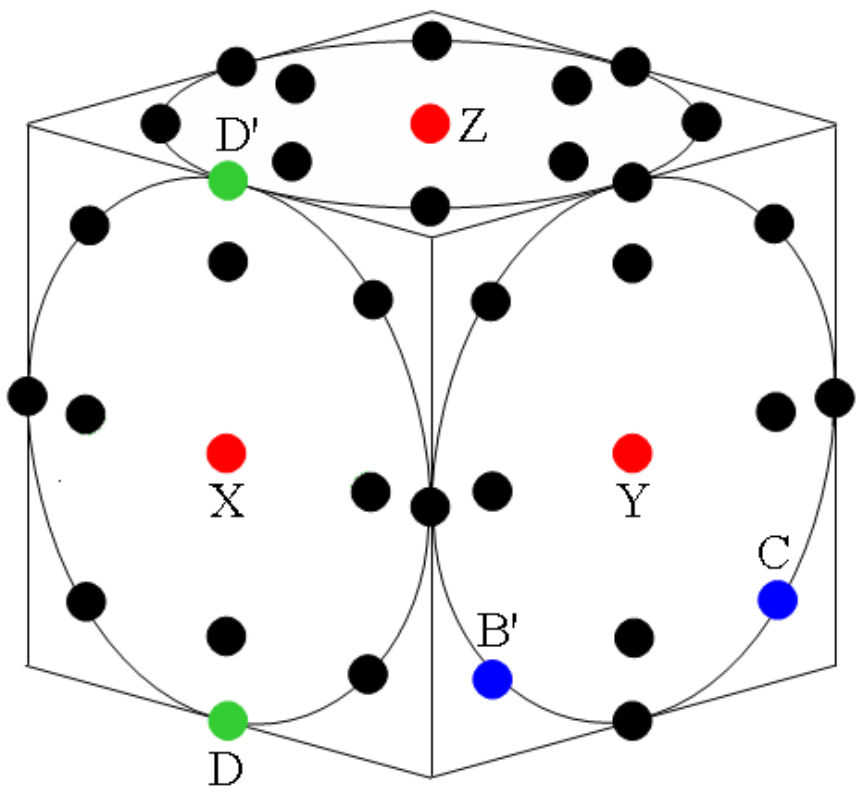

Figure 2. Seven directions in space relevant in estimating the behavior of two bosons.
The red dots together with the center of the cube (not shown) represent the three axis of an orthogonal system. The directions passing through the green dots and the center are mutually perpendicular. The directions passing through the blue dots and the center are auxiliary.

However, if the spin projection of one boson in measured along the direction **Q** and the result is $S_{\mathrm{Q}}^2 = 0$, then, measuring the spin projection of the other boson along a direction **P** with $\mathbf{P} \perp \mathbf{Q}$, the result will be $S_{\mathrm{P}}^2 = 1$. This situation is illustrated in the appendix, where for the directions **Q** and **P** are taken the axis **Z** and an arbitrary direction in the plane x-z, respectively. Expressing the vector $|0\rangle_{\mathrm{Z}}$ in the base {P} one can see that $|0\rangle_{\mathrm{Z}} = |1'\rangle_{\mathrm{P}}$ – see the equations (A2) and (A3). Vice-versa, expressing $|0\rangle_{\mathrm{P}}$ in the base {Z}, the expression contains only the vectors $|1\rangle_{\mathrm{Z}}$ and $|1'\rangle_{\mathrm{Z}}$ – see (A4).

## 3. An experiment with the singlet of spin 1 bosons, and moving frames of coordinates

Two spin 1 bosons, $\mathcal{A}$ and $\mathcal{B}$, are produced in the singlet state (2). The notation is that in each product of states the state of the boson $\mathcal{A}$ is written first. The boson $\mathcal{A}$ flies to the lab of the experimenter Alice, and the boson $\mathcal{B}$ to the lab of the experimenter Bob.

We will consider two frames of coordinates, $\mathcal{F}^1$ and $\mathcal{F}^2$, in movement with respect to one another. Alice's lab will be assumed to be at rest with respect to $\mathcal{F}^1$, and Bob's lab at rest with respect to $\mathcal{F}^2$. The relative velocity of the frames is chosen so that according to the time axis of $\mathcal{F}^1$, by the time the boson $\mathcal{A}$ ends its trip through the setup in Alice's lab, the boson $\mathcal{B}$ only enters Bob's lab – figure 3. Symmetrically, according to the time axis of $\mathcal{F}^2$, by the time the boson $\mathcal{A}$ ends its trip through the setup in Alice's lab, the boson $\mathcal{B}$ only enters Bob's lab – figure 3. Symmetrically according to the time axis of $\mathcal{F}^2$, by the time the boson $\mathcal{B}$ ends its trip through the setup in Bob's lab, the boson $\mathcal{A}$ only enters Alice's lab – figure 4.

Alice measures on her boson the square spin projection along the axis B' and retains only the particles not responding $S_{\mathrm{B'}}^2 = 0$. By her lab-time Bob didn't begin his tests, so that no measurements precedes Alice's measurements. We can write the w-f (2) in the base {B'} and retain only the truncation (non-normalized)

$$|\psi_{\mathrm{tr}_1}\rangle = \frac{1}{\sqrt{3}}\left(|1\rangle_{\mathrm{B'}}|1\rangle_{\mathrm{B'}} - |1'\rangle_{\mathrm{B'}}|1'\rangle_{\mathrm{B'}}\right). \tag{3}$$

In continuation, Alice tests her boson along the direction **D'**. Passing from the base {B'} to the base {D'} for the boson $\mathcal{A}$ – see the transformations (A9) in the appendix – one gets

$$|\psi_{\mathrm{tr}_1}\rangle = \frac{1}{\sqrt{3}}\left(\frac{|1\rangle_{\mathrm{D'}} + \iota|0\rangle_{\mathrm{D'}} - |1'\rangle_{\mathrm{D'}}}{\sqrt{3}}|1\rangle_{\mathrm{B'}} - \frac{\iota|1\rangle_{\mathrm{D'}} + 2|0\rangle_{\mathrm{D'}} + |1'\rangle_{\mathrm{D'}}}{\sqrt{6}}|1'\rangle_{\mathrm{B'}}\right). \tag{4}$$

One can see that one of the outcome of this measurement is $S_{\mathrm{D'}}^2 = 0$. The measurement of $S_{\mathrm{D'}}^2$ collapses the w-f onto a product of independent states of the two particles. Retaining for Alice's boson only the result $S_{\mathrm{D'}}^2 = 0$ there remains

$$|\psi_{\mathrm{tr}_2}\rangle = \frac{1}{3}|0\rangle_{\mathrm{D'}}\left(\iota|1\rangle_{\mathrm{B'}} - \sqrt{2}\,|1'\rangle_{\mathrm{B'}}\right). \tag{5}$$

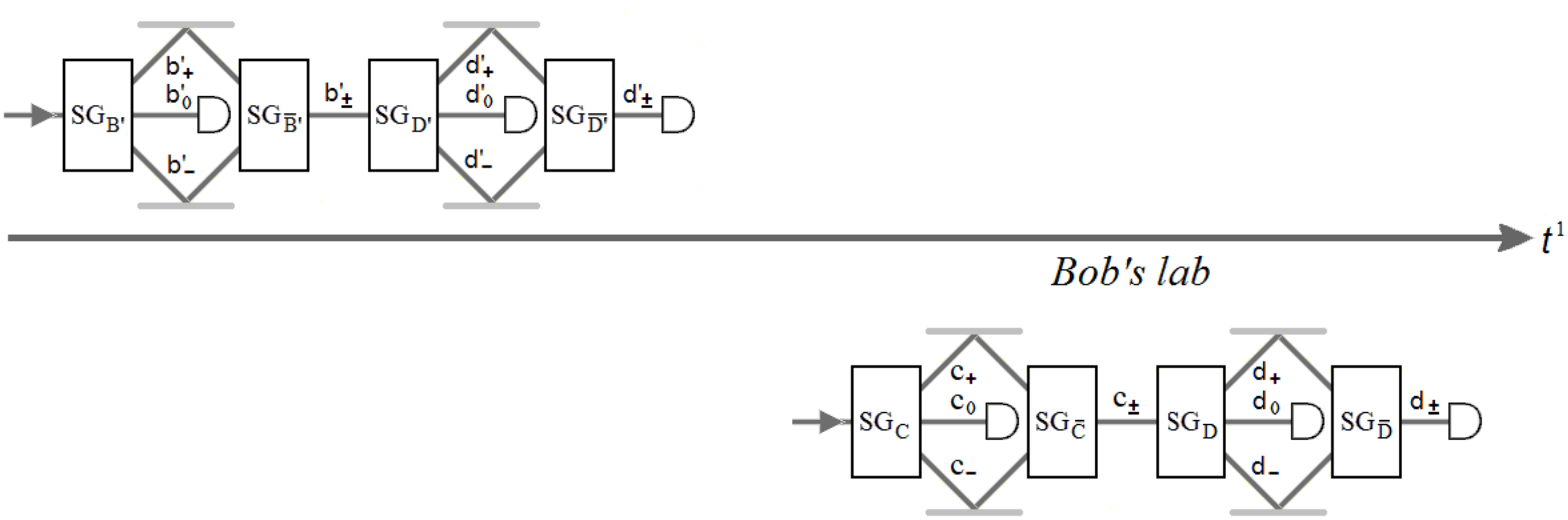

Figure 3. The order of the measurements according to the time-axis of Alice's station.
Alice and Bob pass their bosons through a series of Stern-Gerlach apparatuses. The apparatuses labeled with an upper bar on the subscript merge the beams exiting the apparatuses labeled without upper bar on the subscript. By the time axis of Alice's lab the boson $\mathscr{A}$ exits Alice's lab setup before the boson $\mathscr{B}$ enters Bob's lab setup.

After a while, Bob does the test of his particle along the direction C and retains only the result $S_C^2 = 1$. Passing in (5) from the base {B'} to the base {C} – the transformations (A11) in the appendix – there results

$$|\psi_{tr_3}\rangle = \frac{1}{9}|0\rangle_{D'}\left(\iota\sqrt{2}|1\rangle_C - |1'\rangle_C\right). \quad (6)$$

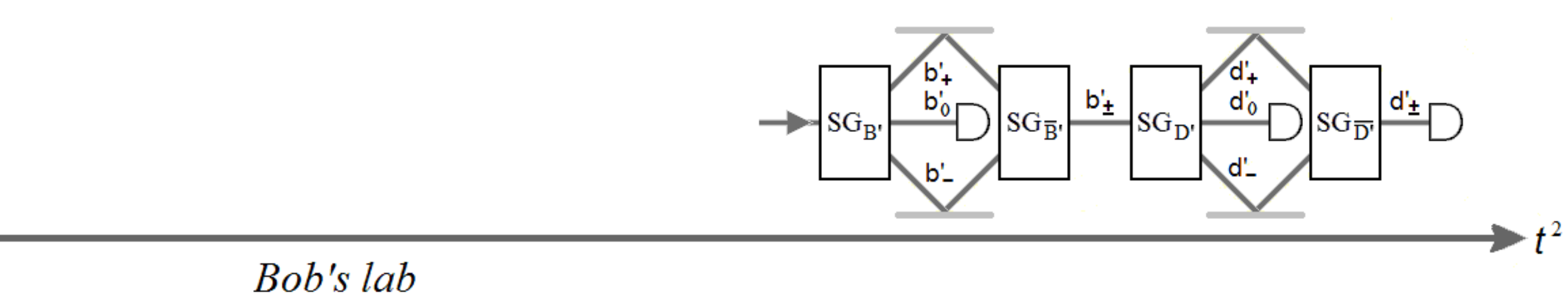

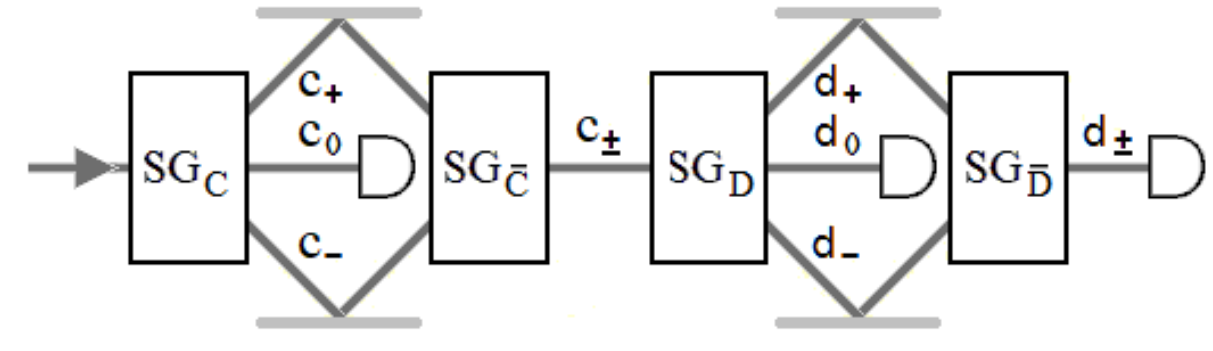

Figure 4. The order of the measurements according to the time-axis of Bob's station.
By the time axis of Bob's lab the boson $\mathscr{B}$ exits Bob's lab setup before the boson $\mathscr{A}$ enters Alice's lab setup.

As Bob measures further the observable $S^2_{\mathrm{D}}$ we have to pass from the base {C} to {D} – the transformations (A10) in the appendix. We are interested in the result $S^2_{\mathrm{D}} = 0$, so, we retain only the vector $|0\rangle_{\mathrm{D}}$,

$$|\psi_{\mathrm{tr}_4}\rangle = \frac{2\sqrt{2}}{9\sqrt{3}}|0\rangle_{\mathrm{D'}}|0\rangle_{\mathrm{D}}. \qquad (7)$$

Therefore, the sequence of measurements as seen according to the frame $\mathcal{F}^1$ leads to a non-zero probability of obtaining $S^2_{\mathrm{D'}} = S^2_{\mathrm{D}} = 0$.

The same final situation appears if judging according to the frame $\mathcal{F}^2$. We can express the singlet in the base {C} and retain only the eigenvectors of $S^2_{\mathrm{C}} = 1$,

$$|\psi_{\mathrm{tr}_5}\rangle = \frac{1}{\sqrt{3}}\left(|1\rangle_{\mathrm{C}}|1\rangle_{\mathrm{C}} - |1'\rangle_{\mathrm{C}}|1'\rangle_{\mathrm{C}}\right). \qquad (8)$$

Passing to the base {D} for Bob's boson, and retaining only the eigenvector of $S^2_{\mathrm{D}} = 0$ – see the transformation (A10) in the appendix – the w-f is reduced to a product of independent states for the two bosons,

$$|\psi_{\mathrm{tr}_6}\rangle = \frac{1}{3}\left(-\iota|1\rangle_{\mathrm{C}} + \sqrt{2}|1'\rangle_{\mathrm{C}}\right)|0\rangle_{\mathrm{D}}. \qquad (9)$$

After a while according to the frame $\mathcal{F}^2$, Alice does her measurements. She begins with the measurement of $S^2_{\mathrm{B'}}$ and retains the result $S^2_{\mathrm{B'}} = 0$. Then, she performs a second measurement, of $S^2_{\mathrm{D'}}$, and retains only $S^2_{\mathrm{D'}} = 0$. After some lengthy, though simple calculus, involving the transformation from the base {C} to {B'} – (A12) in the appendix – and from the base {B'} to {D'} – (A9) in the appendix – there results again a nonzero amplitude of probability for $|0\rangle_{\mathrm{D'}}|0\rangle_{\mathrm{D}}$.

## 4. Nonlocal hidden variables

We make the following hypothesis about the properties of the NLHV:

1. In each trial of the experiment the value of the NLHV is stable all along the trial.
2. The value of the NLHV is well defined and unique in all the space.
3. The NLHV is present in the apparatus and determines the result of any possible test, no matter whether it is performed or not.

According to the feature 4 the NLHV value determines the result of all the four observables, $S^2_{\mathrm{B'}}$ and $S^2_{\mathrm{D'}}$ for the boson $\mathcal{A}$, $S^2_{\mathrm{C}}$ and $S^2_{\mathrm{D}}$ for the boson $\mathcal{B}$, be they measured or not yet measured. In base of the results of the previous section, a set of values $\Lambda$ should exist, that determine $S^2_{\mathrm{B'}} = S^2_{\mathrm{C}} = 1$ and $S^2_{\mathrm{D'}} = S^2_{\mathrm{D}} = 0$.

However, a problem appears about the results $S^2_{D'} = S^2_D = 0$.

The direction **D'** is perpendicular on **B'** – it can be simply checked in (A5) and (A7) that the vectors $|0\rangle_{B'}$ and $|0\rangle_{D'}$ are orthogonal to one another, i.e. $S^2_{B'} = S^2_{D'} = 0$ is impossible. In consequence, as explained in the previous section, the operators $\hat{S}^2_{B'}$ and $\hat{S}^2_{D'}$ commute and could be measured in the opposite order, i.e. first $\hat{S}^2_{D'}$ and then $\hat{S}^2_{B'}$, with the same results, $S^2_{D'} = 0$ and $S^2_{B'} = 1$.

An analogous situation appears for the operators $\hat{S}^2_C$ and $\hat{S}^2_D$. The direction **D** is perpendicular on **C** – check that the vectors $|0\rangle_C$ and $|0\rangle_D$ in (A6) and (A8) are orthogonal to one another – so that $\hat{S}^2_C$ and $\hat{S}^2_D$ commute. So, they could have been measured in opposite order, first $\hat{S}^2_D$ and second $\hat{S}^2_C$ with the same results, $S^2_D = 0$ and $S^2_C = 1$.

In all, the first measurements performed by Alice and Bob could be of the operators $S^2_{D'}$ and $S^2_D$. However, one can see on the figure 2 that $\mathbf{D} \perp \mathbf{D'}$. The propriety (1) mentioned in the section 2 says that for such vectors, if $S^2_D = 0$, $S^2_{D'}$ must be equal to 1, or vice-versa.

Therefore, a contradiction arose. The question is, which values can take the NLHV? Shall it take value that produce $S^2_{D'} = S^2_D = 0$ – the result obtained in the previous section? Shall it take values that produce $S^2_{D'} = 0$ and $S^2_D = 0$, or vice-versa, according the impossibility revealed above? If no value of the NLHV produces $S^2_{D'} = S^2_D = 0$, the NLHV excludes the probability of obtaining this result proved in the former section as nonzero.

## 4. Conclusions

It was proved in this text that the hypothesis of NLHV runs into an impasse. Then, what remains to do is to calculate for each experiment, step by step, the possible outcomes according to the quantum formalism, and not assuming a priori established values for all the involved results.

## Appendix

An arbitrary direction **Q** is defined with respect to an orthogonal triplet of axes, **X**, **Y**, **Z** – see figure 1 – by means of two angles, $\theta$ between the directions **Q** and **Z**, and $\varphi$ between the projection of **Q** on the *x*-*y* plane and the direction **X**. The general base {Q} of eigenvectors of the operator $\hat{S}^2_Q$ is:

$$|1\rangle_Q = \frac{1}{\sqrt{2}}\begin{pmatrix} e^{-\iota\varphi} \\ 0 \\ e^{\iota\varphi} \end{pmatrix}, \quad |0\rangle_Q = \begin{pmatrix} -2^{-1/2}\sin\theta\, e^{-\iota\varphi} \\ \cos\theta \\ 2^{-1/2}\sin\theta\, e^{\iota\varphi} \end{pmatrix}, \quad |1'\rangle_Q = \begin{pmatrix} 2^{-1/2}\cos\theta\, e^{-\iota\varphi} \\ \sin\theta \\ -2^{-1/2}\cos\theta\, e^{\iota\varphi} \end{pmatrix}, \tag{A1}$$

see more details in [22]. For the direction **Z**, defined by $\theta = 0$ and $\varphi = 0$ one gets the base {Z}

$$|1\rangle_Z = \frac{1}{\sqrt{2}}\begin{pmatrix}1\\0\\1\end{pmatrix}, \qquad |0\rangle_Z = \frac{1}{\sqrt{2}}\begin{pmatrix}0\\1\\0\end{pmatrix}, \qquad |1'\rangle_Z = \begin{pmatrix}1\\0\\-1\end{pmatrix}, \tag{A2}$$

For any direction **P** in the plane perpendicular to **Z** one obtains the rotated base

$$|1\rangle_P = \frac{1}{\sqrt{2}}\begin{pmatrix}e^{-\iota\varphi}\\0\\e^{\iota\varphi}\end{pmatrix}, \qquad |0\rangle_P = \frac{-1}{\sqrt{2}}\begin{pmatrix}e^{-\iota\varphi}\\0\\-e^{\iota\varphi}\end{pmatrix}, \qquad |1'\rangle_P = \begin{pmatrix}0\\1\\0\end{pmatrix}, \tag{A3}$$

Expressing the base {P} in the base {Z} one gets

$$\begin{aligned} |1\rangle_P &= \frac{1}{\sqrt{2}}\left(\frac{e^{-\iota\varphi}+e^{\iota\varphi}}{\sqrt{2}}|1\rangle_Z + \frac{e^{-\iota\varphi}-e^{\iota\varphi}}{\sqrt{2}}|1'\rangle_Z\right) \\ |0\rangle_P &= \frac{1}{\sqrt{2}}\left(\frac{e^{-\iota\varphi}-e^{\iota\varphi}}{\sqrt{2}}|1\rangle_Z + \frac{e^{-\iota\varphi}+e^{\iota\varphi}}{\sqrt{2}}|1'\rangle_Z\right). \\ |1'\rangle_P &= |1'\rangle_Z \end{aligned} \tag{A4}$$

The directions **B'** and **C** make with the axis **Z** the angle $\theta = \pi/2 + \pi/6$. Their projections on the plane *x*-*y* are symmetrical with respect to the axis **Y**, so, they make with the axis **X** the angles $\varphi = \pi/2 - \beta$ and $\varphi = \pi/2 + \beta$, respectively, $\sin\beta = 1/\sqrt{3}$. Thus, the direction **B'** is defined as $\mathrm{B'} = (-½, \sqrt{3}/2\sin\beta, \sqrt{3}/2\cos\beta)$, while **C** is defined as $\mathbf{C} = (-½, -\sqrt{3}/2\sin\beta, \sqrt{3}/2\cos\beta)$. Introducing in (2) the angles $\theta$ and $\varphi$, the vectors of the base {B'} are

$$|1\rangle_{B'} = \frac{\iota}{\sqrt{2}}\begin{pmatrix}-e^{\iota\beta}\\0\\e^{-\iota\beta}\end{pmatrix}, \qquad |0\rangle_{B'} = \frac{\iota\sqrt{3}}{2\sqrt{2}}\begin{pmatrix}e^{\iota\beta}\\\iota\sqrt{2/3}\\e^{-\iota\beta}\end{pmatrix}, \qquad |1'\rangle_{B'} = \frac{\iota}{2\sqrt{2}}\begin{pmatrix}e^{\iota\beta}\\-\iota\sqrt{6}\\e^{-\iota\beta}\end{pmatrix}. \tag{A5}$$

The vectors of the base {C} can be obtained from those of the base {B'} by replacing $\beta$ with $-\beta$,

$$|1\rangle_{C} = \frac{\iota}{\sqrt{2}}\begin{pmatrix}-e^{-\iota\beta}\\0\\e^{\iota\beta}\end{pmatrix}, \qquad |0\rangle_{C} = \frac{\iota\sqrt{3}}{2\sqrt{2}}\begin{pmatrix}e^{-\iota\beta}\\\iota\sqrt{2/3}\\e^{\iota\beta}\end{pmatrix}, \qquad |1'\rangle_{C} = \frac{\iota}{2\sqrt{2}}\begin{pmatrix}e^{-\iota\beta}\\-\iota\sqrt{6}\\e^{\iota\beta}\end{pmatrix}. \tag{A6}$$

The directions **D'** and **D** are perpendicular on one another. They make with the axis **Z** the angles $\theta = \pi/4$, respectively $\theta = 3\pi/4$ and their projections on the plane *x*-*y* fall on the axis **X**, so that $\varphi = 0$. In consequence, they are defined by $\mathbf{D'} = (1/\sqrt{2}, 1/\sqrt{2}, 0)$, respectively $\mathbf{D} = (-1/\sqrt{2}, 1/\sqrt{2}, 0)$, and the vectors of the bases {D} and {D'} are

$$|1\rangle_{D'} = \frac{1}{\sqrt{2}}\begin{pmatrix}1\\0\\1\end{pmatrix}, \qquad |0\rangle_{D'} = \frac{1}{2}\begin{pmatrix}-1\\\sqrt{2}\\1\end{pmatrix}, \qquad |1'\rangle_{D'} = \frac{1}{2}\begin{pmatrix}1\\\sqrt{2}\\-1\end{pmatrix}, \tag{A7}$$

$$|1\rangle_{D} = \frac{1}{\sqrt{2}}\begin{pmatrix}1\\0\\1\end{pmatrix}, \qquad |0\rangle_{D} = -\frac{1}{2}\begin{pmatrix}1\\\sqrt{2}\\-1\end{pmatrix}, \qquad |1'\rangle_{D} = \frac{1}{2}\begin{pmatrix}-1\\\sqrt{2}\\1\end{pmatrix}, \tag{A8}$$

For passing in (3) from the base {B'} to the base {D'} we have the following transformations:

$$|1\rangle_{B'} = \frac{|1\rangle_{D'} + \iota|0\rangle_{D'} - |1'\rangle_{D'}}{\sqrt{3}}, \qquad |1'\rangle_{B'} = \frac{\iota|1\rangle_{D'} + 2|0\rangle_{D'} + |1'\rangle_{D'}}{\sqrt{6}}. \tag{A9}$$

$$_{D'}\langle 1|1\rangle_{B'} = \sin\beta = \frac{1}{\sqrt{3}}, \qquad _{D'}\langle 0|1\rangle_{B'} = \frac{\iota}{\sqrt{2}}\cos\beta = \frac{\iota}{\sqrt{3}}, \qquad _{D'}\langle 1'|1\rangle_{B'} = -\frac{1}{\sqrt{2}}\cos\beta = -\frac{1}{\sqrt{3}},$$

$$_{D'}\langle 1|0\rangle_{B'} = \frac{\iota\sqrt{3}}{2}\cos\beta = \frac{\iota\sqrt{2}}{2}, \qquad _{D'}\langle 0|0\rangle_{B'} = 0, \qquad _{D'}\langle 1'|0\rangle_{B'} = \frac{\iota\sqrt{3}}{2\sqrt{2}}(\iota\sin\beta + \iota\frac{1}{\sqrt{3}}) = -\frac{\sqrt{2}}{2}$$

$$_{D'}\langle 1|1'\rangle_{B'} = \frac{\iota}{2}\cos\beta = \frac{\iota}{\sqrt{6}}, \qquad _{D'}\langle 0|1'\rangle_{B'} = \frac{1}{2\sqrt{2}}(\sin\beta + \sqrt{3}) = \frac{2}{\sqrt{6}}, \qquad _{D'}\langle 1'|1'\rangle_{B'} = \frac{1}{2\sqrt{2}}(-\sin\beta + \sqrt{3}) = \frac{1}{\sqrt{6}}.$$

For passing in (5) from the base {C} to the base {D} we have the following transformations:

$$|1\rangle_{C} = \frac{|1\rangle_{D} - \iota|0\rangle_{D} - \iota|1'\rangle_{D}}{\sqrt{3}}, \qquad |1'\rangle_{C} = \frac{\iota|1\rangle_{D} - 2|0\rangle_{D} + |1'\rangle_{D}}{\sqrt{6}}. \tag{A10}$$

$$_{D}\langle 1|1\rangle_{C} = \sin\beta = \frac{1}{\sqrt{3}}, \qquad _{D'}\langle 0|1\rangle_{B'} = -\frac{\iota}{\sqrt{2}}\cos\beta = -\frac{\iota}{\sqrt{3}}, \qquad _{D'}\langle 1'|1\rangle_{B'} = -\frac{\iota}{\sqrt{2}}\cos\beta = -\frac{\iota}{\sqrt{3}},$$

$$_{D'}\langle 1|1'\rangle_{C} = \frac{\iota}{2}\cos\beta = \frac{\iota}{\sqrt{6}}, \qquad _{D'}\langle 0|1'\rangle_{C} = \frac{1}{2\sqrt{2}}(\sin\beta - \sqrt{3}) = -\frac{2}{\sqrt{6}}, \qquad _{D'}\langle 1'|1'\rangle_{C} = \frac{1}{2\sqrt{2}}(-\sin\beta + \sqrt{3}) = \frac{1}{\sqrt{6}}.$$

The expressions of the vectors $|1\rangle_{B'}$ and $|1'\rangle_{B'}$ in the base {C} are

$$|1\rangle_{B'} = \frac{|1\rangle_{C} - \iota\sqrt{6}|0\rangle_{C} - \iota\sqrt{2}|1'\rangle_{C}}{3}, \qquad |1'\rangle_{B'} = \frac{-\iota 2\sqrt{2}|1\rangle_{C} - \sqrt{3}|0\rangle_{C} + 5|1'\rangle_{C}}{6}. \tag{A11}$$

$$_{C}\langle 1|1\rangle_{B'} = \cos(2\beta) = 1/3, \qquad _{C}\langle 0|1\rangle_{B'} = -\frac{\iota\sqrt{3}}{2}\sin(2\beta) = -\iota\frac{\sqrt{2}}{\sqrt{3}}, \qquad _{C}\langle 1'|1\rangle_{B'} = -\frac{\iota}{2}\sin(2\beta) = -\iota\frac{\sqrt{2}}{3}.$$

$$_{C}\langle 1/1'\rangle_{B'} = -\iota\frac{\sin(2\beta)}{2} = -\iota\frac{2\sqrt{2}}{6}, \qquad _{C}\langle 0/1'\rangle_{B'} = \frac{\sqrt{3}}{4}[\cos(2\beta)-1] = -\frac{\sqrt{3}}{6}, \qquad _{C}\langle 1'/1'\rangle_{B'} = \frac{1}{4}[\cos(2\beta)+3] = \frac{5}{6}.$$

The expressions of the vectors $/1\rangle_C$ and $/1'\rangle_C$ in the base $\{B'\}$ are

$$/1\rangle_C = \frac{1}{3}/1\rangle_{B'} + \iota\frac{\sqrt{2}}{3}/1'\rangle_{B'}, \qquad /1'\rangle_C = \iota\frac{\sqrt{2}}{3}/1\rangle_{B'} + \frac{5}{6}/1'\rangle_{B'}. \tag{A12}$$

$$_{B'}\langle 1/1\rangle_C = \cos(2\beta) = 1/3, \qquad _{B'}\langle 1'/1\rangle_{C'} = \iota\frac{\sin(2\beta)}{2} = \iota\frac{\sqrt{2}}{3},$$

$$_{B'}\langle 1/1'\rangle_C = \frac{\iota}{2}\sin(2\beta) = \iota\frac{\sqrt{2}}{3}, \qquad _{B'}\langle 1'/1'\rangle_C = \frac{1}{4}[\cos(2\beta)+3] = \frac{5}{6}.$$

---